\documentclass[useAMS,usenatbib]{mn2e}
\usepackage{natbib}
\usepackage{color,graphicx} 
\usepackage{amsmath}        
\usepackage{amsfonts}       
\usepackage{amssymb}        
\usepackage{wasysym}        

\author[Keane et al.]{E. F.~Keane$^{\dagger1}$, D. A. Ludovici$^{3}$,
R. P. Eatough$^1$, M.~Kramer$^{1,2}$, A. G. Lyne$^1$, \newauthor \ 
M. A. McLaughlin$^3$ \& B. W. Stappers$^1$ \\ $^{\dagger}$ Enquiries to:
ekean@jb.man.ac.uk \\ $^1$ University of Manchester, Jodrell Bank
Centre for Astrophysics, Alan Turing Building, Oxford Road, Manchester
M13 9PL, UK. \\ $^2$ Max Planck Institut f\"{u}r Radioastronomie, Auf
dem H\"{u}gel 69, 53121 Bonn, Germany. \\ $^3$ Department of Physics,
West Virginia University, Morgantown, WV 26506, USA.}

\date{18 June 2009} 

\title[RRAT Searches of the PMPS]{Further Searches for RRATs in the
Parkes Multi-Beam Pulsar Survey}

\begin{document}

\maketitle

\begin{abstract}
  We describe the steps involved in performing searches for sources of
  transient radio emission such as Rotating Radio Transients (RRATs),
  and present 10 new transient radio sources discovered in a
  re-analysis of the Parkes Multi-beam Pulsar Survey.  Followup
  observations of each new source as well as one previously known
  source are also presented. The new sources suggest that the
  population of transient radio-emitting neutron stars, and hence the
  neutron star population in general, may be even larger than
  initially predicted. We highlight the importance of radio frequency
  interference excision for single-pulse searches. Also, we discuss
  some interesting properties of individual sources and consider the
  difficulties involved in precisely defining a RRAT and determining where they
  fit in with the other known classes of neutron stars.
\end{abstract}


\section{Introduction}
Rotating Radio Transients (RRATs from herein) were discovered in a
search of archival data from the Parkes Multi-beam Pulsar Survey (PMPS
from herein) by~\citet{mll+06}. These sources exhibit infrequent,
short, relatively bright bursts of radio emission. The pulses have
peak flux densities (at 1.4 GHz) ranging from $\sim100$ mJy up to
$\sim10$ Jy and pulse widths in the range $2 - 30$ ms. As we detect
pulses only occasionally from RRATs they are generally not detected in
Fourier domain searches (although see discussion in section 3) but
rather by searching for individual bright
pulses~\citep{mc03,cm03}. Periodicities can be determined by
factorising the difference between each pair of pulse arrival times
(as discussed in Section 2). The underlying periodicities are in the
range $0.7 - 6.7$ s. These periodicities are believed to be spin
periods and thus RRATs seem to be a new population of neutron
stars. 
Further evidence for this comes from X-ray observations of RRAT
J1819$-$1458 (the most prolific, brightest and thus well studied
source) which have revealed thermal emission consistent with that
expected from cooling neutron stars as well as an X-ray period
identical to that determined from radio
observations~\citep{rbg+06,mrg+07}.

We see these millisecond bursts as infrequently as once per hour and
up to as often as every few minutes. The long periods observed are
more comparable to the magnetars~\citep{wt_06} and the Isolated
Neutron Stars (INSs, aka XDINSs,~\citet{k08}), than the radio pulsars
and the relationship between these populations is an open
question. The narrow pulse widths are comparable to individual pulses
from radio pulsars. Coupled to their longer periods the duty cycles
are apparently much smaller, but including the entire phase range
wherein we see emission gives duty cycles comparable to integrated
pulsar profiles, e.g J1819$-$1458 shows emission over $\sim2.9\%$
phase range~\citep{glitch_paper}. There have been many suggested
explanations for the `bursty' RRAT behaviour. Some assign it to
detection issues with others favouring intrinsically transient
emission. \citet{wsrw06} support the former notion and suggest that
the RRAT emission may be due to these sources being more-distant
analogues of PSR B0656$-$14, i.e. pulsars whose regular emission may
be below our detection limit but who show detectable giant pulses or
have an amplitude distribution with a long tail at high flux
densities. The alternative models view the emission as truly
intermittent with the short duration bursts corresponding to
activation times of external excitation events. This is the case in
the model of Cordes and Shannon (2006)\nocite{cs08} which considers
the re-activation of inactive vacuum gaps due to the presence of
circum-stellar asteroidal material. Similarly it has been proposed
by~\citet{lm07} that RRATs may be surrounded by a radiation belt which
could stimulate transient emission in the magnetosphere. 
behaviour of RRAT

Apart from these unusual emission properties RRATs are also of
interest due to the predicted size of the population, which is thought
to be several times larger than the radio pulsar population~\citep{mll+06}.
Previously~\citet{keanekramer08} discussed the implications of
such a large number of RRATs on the Galactic production rate of
neutron stars. It was argued that the many classes of neutron stars
now known do not seem to be accounted for by the Galactic
core-collapse supernova rate. An evolutionary link between the neutron
star classes might eliminate this birthrate problem. An alternative
explanation would be that the population estimates for RRATs and other
neutron star populations are hugely over-estimated.

The discovery of RRATs has sparked a renewed interest in time-domain
searches for transient radio sources - a relatively unexplored
parameter space~\citep{clm04,lbb+09} which will be revolutionised with
upcoming instruments such as LOFAR~\citep{vs07}, the ATA~(e.g.~van
Leeuwen et al.~2009)\nocite{ata_paper}, the SKA
pathfinders~(e.g.~Johnston et al.~2009)\nocite{jfg09},
FAST~\citep{nwz+06,slk+09} and eventually the
SKA~\citep{ckl+04,sks+09}. For now, addressing these questions
requires improved population estimates and characterisation. In
particular, we need to understand which potential selection effects
are present during the discovery process and how they impact on the
discovered population. With a better understanding of RRAT
characteristics and the potential selection effects an improved
population estimate for RRATs will be possible. Motivated by this we
have performed a full re-processing of the PMPS searching for sources
of single pulses of radio emission. In section 2 below we outline the
steps involved in our re-processing. In Section 3 we discuss in some
detail the observed characteristics of 10 newly discovered sources as
well as the results of follow-up work on one detected, but previously
known, source. For the purposes of sections 2 and 3 we will consider
that for a source to be a RRAT at some frequency that it is necessary
but not sufficient that it is detected more easily in a single pulse
search than in a periodicity search. In section 4 we discuss in more
detail a refined definition of what a RRAT is as well as the
implications for the Galactic neutron star population of these new
discoveries before concluding with an outlook towards surveys with
LOFAR and the SKA.

\section{PMSingle}
In the original RRAT discovery paper~\citep{mll+06} it was estimated
that approximately half of the RRATs visible in the PMPS had been
detected, with the remainder obscured due to the effects of radio
frequency interference (RFI). Recently it has become timely to
re-process the PMPS survey in search of these postulated sources as we
have developed a new and effective RFI mitigation
scheme~\citep{ekl09}. Utilising the Jodrell Bank Pulsar Group's
recently acquired 1400-processor HYDRA super-computer, the whole PMPS
data set was re-processed. Modified versions of the
\textit{Sigproc}\footnote{http://sigproc.sourceforge.net/} processing
tools were used. In the following we refer to this project as the
PMSingle process.

\begin{figure*}  
  \begin{center}
    \includegraphics[trim = 0mm 12mm 0mm 125mm, clip,scale=0.4]{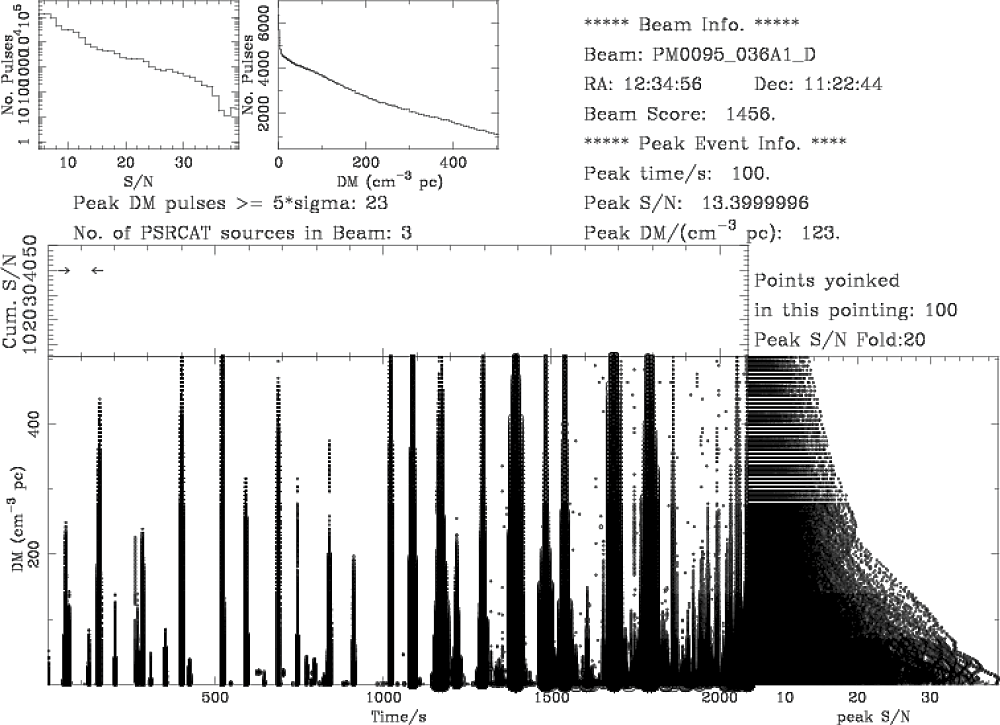} 
    \includegraphics[trim = 0mm 12mm 0mm 128mm, clip,scale=0.4]{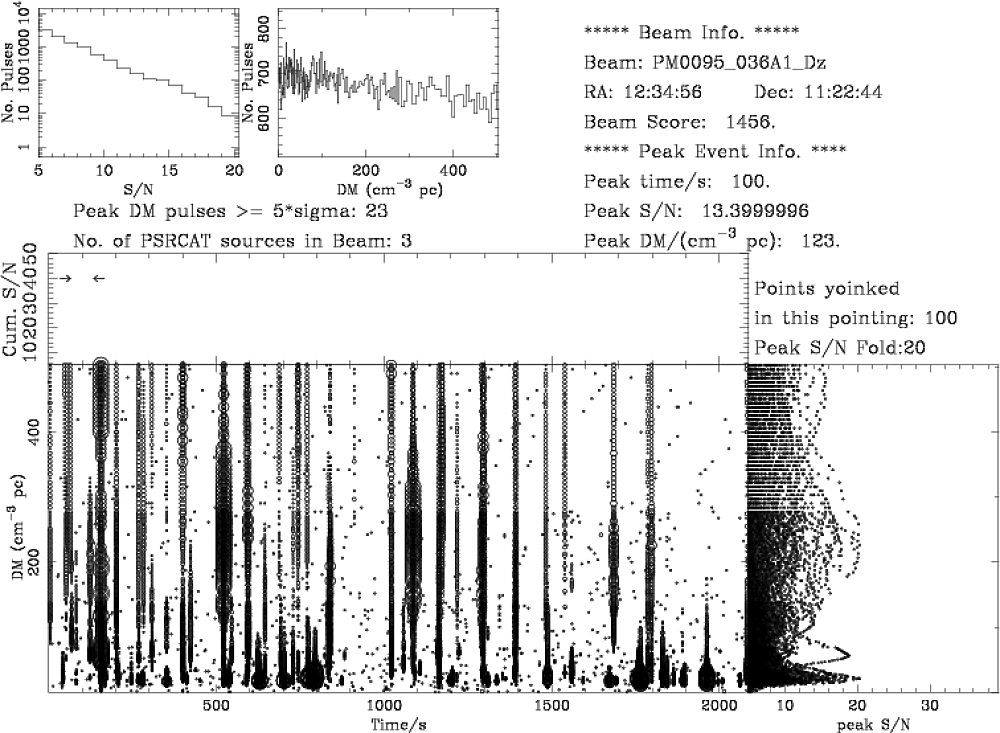}
    \includegraphics[trim = 0mm 0mm 0mm 128mm, clip,scale=0.4]{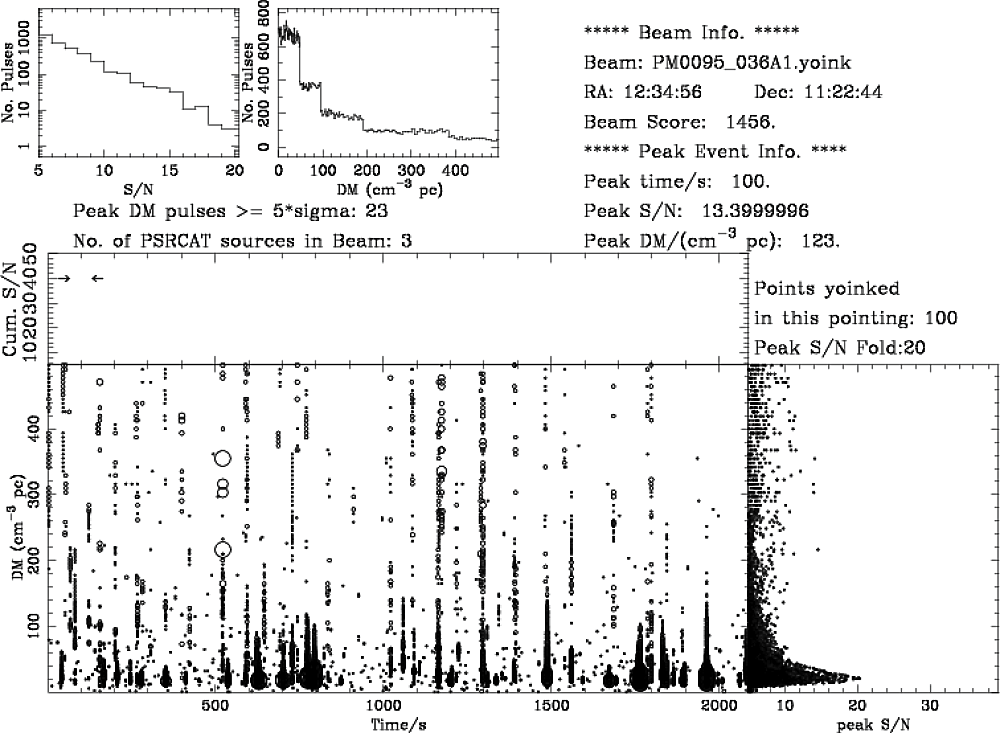} 
  \end{center}      
  \caption{\small{ The ordinate in the large left-hand plots is trial
  DM over the range $0-500$ cm$^{-3}$pc and the abscissa is time over
  a 35-minute PMPS observation. To the right are plots of trial DM
  versus peak signal-to-noise ratio. Significant detected events are
  plotted as circles with radius proportional to signal-to-noise. (a)
  The strong vertical stripes across wide DM ranges are instances of
  extremely strong RFI. Inspection of the plot for the presence of a
  celestial source is impossible due to the presence of this RFI. (b)
  The same beam after the zero-DM filter has been applied (see
  \S2(ii)). We can see that the RFI has not been completely removed,
  especially at higher DMs. However the RFI has been removed much more
  effectively at low DMs and a source at DM$\sim20$ cm$^{-3}$pc is
  beginning to become visible. (c) The same beam after application of
  the zero-DM filter and the removal of multiple-beam events. We can
  see that the diagnostic plot is cleaned up even further and,
  although there is still remnant RFI, it is clear that there is a
  real source in this beam at DM$\sim20$ cm$^{-3}$pc. This is the
  first detection of J1841$-$14 in the PMPS, the lowest DM source
  found so far in the survey.}}
  \vspace{-15pt}
  \label{fig:lowDM_blindness}
\end{figure*}

We note that the presence of RFI will make a search somewhat blind to
sources with low dispersion measure (DM). The DM of a source at a
distance $L$ from Earth is
\begin{equation}
  DM=\int_{0}^{L}n_{\rm{e}}(l)dl,
\end{equation}
where $n_{\rm{e}}(l)$ is the electron density at a distance $l$. RFI
is terrestrial in origin and, not having traversed the inter-stellar
medium, we expect it to have $DM=0$.\footnote{RFI signals emerging
from air-traffic control radar, a particular problem at frequencies
near 1400~MHz, are sometimes observed to also show signals sweeping in
frequency.}  However as RFI signals are typically very strong, in
comparison to the relatively weak astrophysical signals of interest,
they are seen with sufficient residual intensity to mask celestial
signals to high DM values. Figure~\ref{fig:lowDM_blindness} shows an
example of this `low-DM blindness' due to the presence of strong
terrestrial RFI. This of course means that searches can miss real,
low-DM sources. Thus our sensitivity to the nearby Galactic volume may
have been reduced due to the effects of RFI in the initial analysis.

Before proceeding with outlining the PMSingle re-processing steps we
note some survey and data specifications: The survey covered a strip
along the Galactic plane with $|b|<5^o$ and between $l=260^0$ and
$l=50^o$. It consists of $3196\times35$-minute pointings at an
observation frequency of 1.4~GHz using the Parkes 21-cm multi-beam
receiver. This receiver has 13 beams which each receive orthogonal
linear polarisation so that there are $41561$ dual-polarisation beams
in total. An analogue filterbank with $96\times3$ MHz channels was
used with 250-$\mu$s sampling. Once the data were received both
polarisation signals were added to give total intensity (Stokes I) and
the data were 1-bit digitised before recording. Thus the raw data for
each beam can be visualised as a 1-bit digitised data cube (time,
frequency and amplitude). The survey specifics are described in more
detail in~\citet{mlc+01}.

Our PMSingle processing involved the following steps:
\begin{enumerate}
\item {\em Remove all clipping algorithms.} In previous analyses of
the PMPS the raw data have been `clipped', i.e.~the data were read in
48 KB blocks and those blocks wherein the sum of all the bit values
was larger than some (user supplied) threshold above the mean had
their values set to the mean level (half 1s and half 0s in the case of
1-bit data). The motivation for such clipping is that RFI signals are
typically much stronger than real astrophysical signals, so that the
brightest detections are taken to be RFI spikes. This, however, is not
optimal in that it removes signals based on strength and the discovery
of RRATs and pulsars which show strong single pulses show that such
signals may be of astrophysical interest. The threshold used is also
arbitrary and usually determined on a trial-and-error basis. We note
that the strongest pulses with high dispersion (i.e. dispersed over 2
or more blocks) can escape being clipped in error so that low-DM
sources are the most likely to be clipped in this way. We therefore
removed this step from our re-processing.

\item {\em Dedisperse the raw filterbank data using the zero-DM
filter.} We searched for dispersed signals in a DM range of $0-2200$
$\rm{cm^{-3}.pc}$. For the Galactic longitudes covered by the PMPS
this corresponds, at $|b|=0^o$, to typical distances of $\gtrsim40$
kpc at $l=260^0$ and $l=50^o$ to $8.5$ kpc towards the Galactic
centre. The zero-DM filtering technique can be simply stated: for each
time sample $t_{\rm{j}}$ we average all $N_{\rm{chans}}$ frequency
channels and subtract this average from each channel. The value of a
frequency channel $f_{\rm{i}}$ at time $t_{\rm{j}}$ is
$S(f_{\rm{i}},t_{\rm{j}})$ and after applying the zero-DM filter
function will become:
\begin{align}
  S(f_{\rm{i}},t_{\rm{j}})&\rightarrow
  S^{\prime}(f_{\rm{i}},t_{\rm{j}}) \nonumber \\
  &=S(f_{\rm{i}},t_{\rm{j}})-\frac{1}{N_{\rm{chans}}}\sum_{i=1}^{N_{\rm{chans}}}S(f_{\rm{i}},t_{\rm{j}}).
\end{align}
This has the effect of removing short-duration broadband RFI but real
dispersed pulses will be convolved with a particular function. After
applying this filter we dedisperse the data. A dispersed input square
pulse $\sqcap(t,$DM$,w,BW,f_{0})$ at time $t$, with pulse width $w$,
observed with a bandwidth $BW$ at a central frequency $f_0$ will
become:
\begin{align}
  \sqcap&(t,\mbox{\rm DM},w,BW,f_{0})\rightarrow\mathfrak{W}(t,\mbox{\rm DM},w,BW,f_{0})
  \nonumber \\
  &=\sqcap(t,\mbox{\rm DM},w,BW,f_{0})-\bigtriangledown(t,\mbox{\rm DM},w,BW,f_{0}).
\end{align}
The peak of the pulse is reduced with the addition of triangular
`dips' either side of the pulse with the exact shape dependent on the
pulse's DM and width, as well as on $f_{0}$ and $BW$. Examples of
this are given in Figure~\ref{fig:zdm} which shows single pulses
detected from J1841$-$14. We note that the zero-DM filter acts as a
more natural RFI mitigation tool than standard clipping as it removes
signals based on their dispersion properties as opposed to sheer
strength. Details of the zero-DM filter including assumptions, uses
and limitations are discussed in detail in~\citet{ekl09}.
\begin{figure}  
  \begin{center}
    \includegraphics[trim = 22mm 10mm 17mm 4mm, clip, scale=0.32,angle=-90]{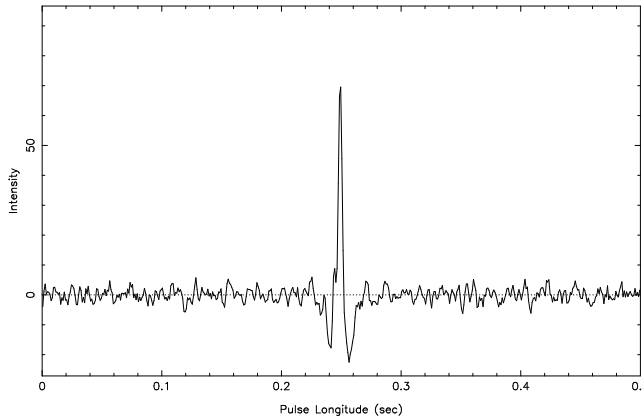}
  \end{center}      
  \caption{\small{An example of a single pulse detected from 
  J1841$-$14. Here we plot 0.5 seconds of the 6.6-second period
  centred on the pulse with intensity plotted in arbitrary
  units. Evident are the dips either side of the pulse which is a
  remnant of the zero-DM filtering process.}}
  \label{fig:zdm}
\end{figure}

\item {\em Search for bright single pulses.} This is an excercise in
matched filtering using box-cars of various sizes. However, instead of
rectangular box-cars the zero-DM filtering means it is optimal to
search with box-cars of the form $\mathfrak{W}(t,DM,w,BW,f_{0})$ for
various trial widths. That this is the optimal strategy is evident
from examining the pulse in Figure~\ref{fig:zdm} which clearly shows
the characteristic dips from the zero-DM filtering procedure. In
PMSingle we search for pulse widths as narrow as 250 $\rm{\mu}$s to as
wide as 128 ms at the lowest DMs. As we increase the DM trial value we
get dispersive smearing of pulses to widths much longer than their
intrinsic widths so at these DMs we search for even wider pulses. The
widest pulse widths searched for are a factor of 16 larger than in the
original PMPS single-pulse search analysis of~\citet{mcl_iau}.


\item {\em Perform a beam comparison to remove multi-beam events.} The
zero-DM filtering is effective at removing short-duration broad-band
RFI. The more persistant or narrow-band that RFI is the less likely it
will be completely removed. However as the PMPS used a 13-beam
receiver we have extra information to help with RFI mitigation. Pulsar
signals are very weak and typically are seen in only one beam. The
strongest pulsars (e.g.~Vela) can be seen in a few beams but normally
no more than three. Even the extremely bright 30-Jy 5-ms single burst
(from an unknown source possibly at a cosmological distance) reported
by~\citet{lbm+07}, was seen in just three beams. We can thus apply a
rejection criteria for detected events like: for each detected event -
$(DM,time)$ point, if we have detections in the range
$(DM\pm\varepsilon_{DM}, time\pm\varepsilon_{time})$ for, e.g.
$\geq5$, beams in that pointing then ignore this detection as it is
most likely RFI. We conservatively took $\varepsilon$ to be one bin in
each case (i.e. one DM trial step and one time sample step). In
addition, for all beams, a check is made against the known pulsars
(from the ATNF pulsar
database\footnote{http:/www.atnf.csiro.au/research/pulsar/psrcat/},~\citet{mhth05})
which fall within the beam.

\item {\em Produce diagnostic plots for inspection and
classification.} Along the lines of~\citet{cm03} a series of
diagnostic plots are created for each beam. An example of this is
shown in Figure~\ref{fig:search_plots}. The plots include beam
information (beam and pointing number, sky position etc.) as well as
information on the number of multiple beam detections which were
removed. Each beam was inspected and classified as containing either
noise, known pulsars, known RRATs or new candidates - divided into
Classes 1, 2 and 3. Examples of each of these classes are given in
Figure~\ref{fig:cand_classes}. Class 1 candidates are all thought to
be real sources, either yet-to-be confirmed RRATs or known pulsars
detected in the telescope's far-side-lobes. Class 3 candidates are
weak and no confirmations are expected. Class 2 sources are
intermediate between these classes. Beams could also be classified as
being too adversely affected with remnant RFI (not removed by zero-DM
filtering or beam comparisons) so as to make inspection impossible. In
these beams a real source, unless it were very strong, would not have
been detectable. The results of the classifications are given in
Table~\ref{tab:pmsingle_classifications}.
\begin{figure*}  
  \begin{center}
    \includegraphics[scale=0.6,angle=-90]{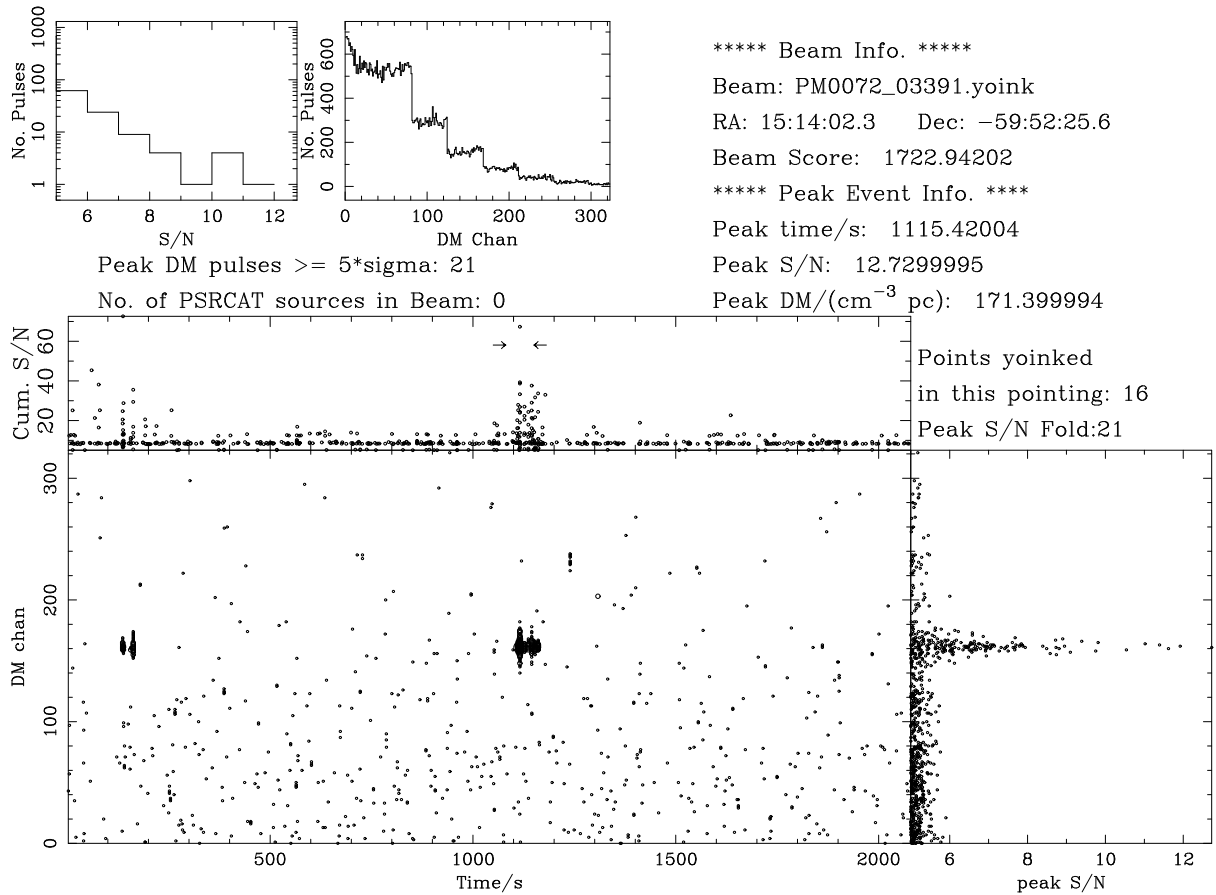}
  \end{center}      
  \caption{\small{An example of the diagnostic plots. This plot shows
  the original detection of J1514-59. Two clumps of pulses are evident
  at $\sim150$ s and at $\sim1100$ s.}}
  \label{fig:search_plots}
\end{figure*}
\begin{figure*}  
  \begin{center}
    \includegraphics[scale=0.5]{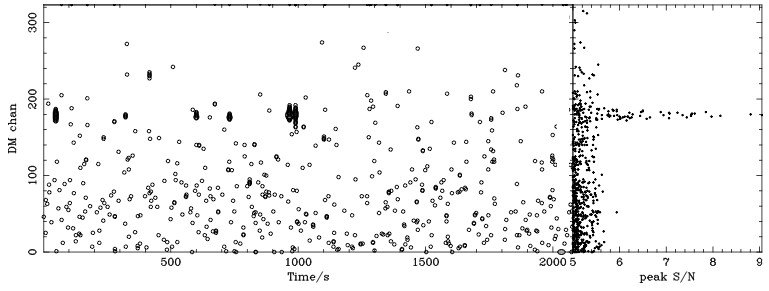}
    \includegraphics[scale=0.42]{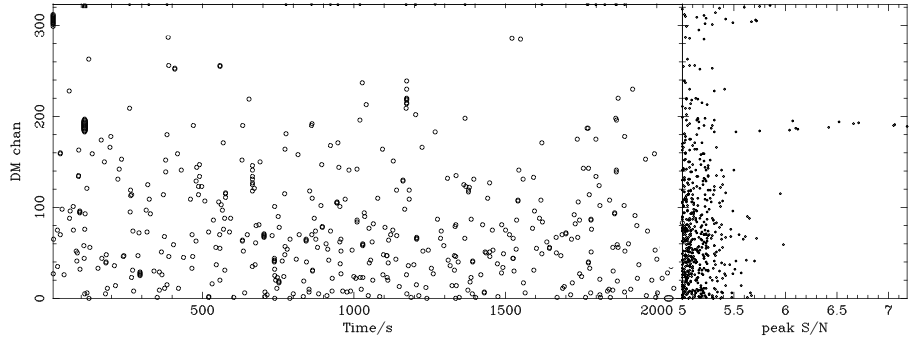}
    \includegraphics[scale=0.5]{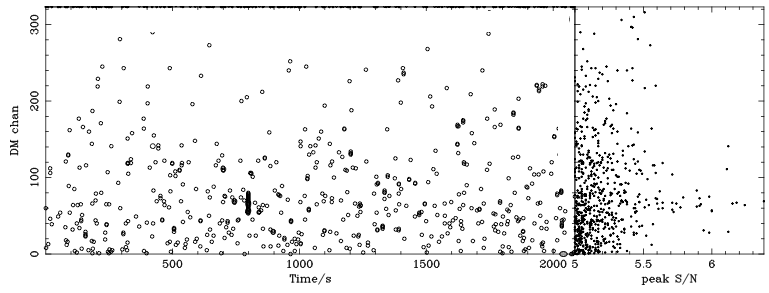}
  \end{center}      
  \caption{\small{(Top) An as yet unconfirmed class 1 candidate with
  several strong pulses at constant DM which appears to be a real
  celestial source; (Middle) A class 2 source showing one significant
  pulse; (Bottom) A typical Class 3 candidates showing a weak single
  pulse. Class 3 sources are the least significant with no
  confirmations expected, especially due to the impractical nature of
  following up such weak sources.}}
  \label{fig:cand_classes}
\end{figure*}

\begin{table}
  \begin{center}
    \caption{The classifications of the PMPS beams in the PMSingle
    single-pulse search. For the known sources the number of unique
    sources detected are given in parentheses.}
    \begin{tabular}{|l|c}
      \hline\hline Classification & $N_{\rm{detections}}$ \\
      \hline Candidate:Class 1 & 162 \\
      Candidate:Class 2 & 204 \\
      Candidate:Class 3 & 319 \\
      Known PSR & 606 (300)\\
      Known RRAT & 13 (11) \\
      Noise & 27493 \\
      Noise + some remnant RFI & 12061 \\
      Large remnant RFI & 693 \\
      \hline
    \end{tabular}
  \label{tab:pmsingle_classifications}
  \end{center}
\end{table}

\item {\em Cross-check with known sources.}
For each candidate we confirm that it is not a known pulsar (or
RRAT). Even if there is no pulsar within the beam the pulses could
still be from a known (strong) pulsar perhaps several beams away on
the sky (e.g. PSRs B0835$-$41, B0833$-$45 and B1601$-$52 are detected
many times like this). To do this we can compare the position (with a
larger tolerance) and DM of the candidate to that of known sources.

\item {\em Determine a period for the candidate.} As RRATs are not
seen in Fast Fourier Transform (FFT) searches (although see discussion
in Section 3) we use a method of factorising the pulse time of arrival
(TOA) differences to determine the period. For $N$ TOAs there are at
most\footnote{TOA differences can only be used if the TOAs are from
the same observation as we quickly lose phase coherence of TOAs.}
$(N/2)(N-1)$ unique TOA differences which we can use. For some small
increment we step through a large range of trial periods. The correct
period will fit all of the TOA differences with a small error and rms
residual. Harmonics of the true period will also match many of the TOA
differences but at a less significant level. If the most significantly
matching period does not match all of the TOA differences then
progressively removing the TOAs with the largest uncertainty will
increase the significance of the match onto the true period. As a rule
of thumb $\sim8$ pulses (i.e. 28 TOA differences) in an observation
allow a reliable period estimate, i.e. at the correct harmonic. We
note that for a TOA difference to be usable in this method both pulses
must not be far-separated so as not to lose coherence, e.g. 8 pulses
in a single observation yields 28 usable TOA differences for period
determination, but 2 separated observations each of 4 pulses yields
just 12 TOA differences. Some example output plots from this method
are shown in Figure~\ref{fig:1841_getper}. If a period determination
is possible, it can be used with position and DM to further confirm
whether or not the candidate is a previously known source. We note one
difficulty with this method is that it is possible that the range of
emitting pulse longitudes is wide (e.g.~for an aligned rotator) so
that sharp peaks like those shown in Figure~\ref{fig:1841_getper} will
be smeared out across a wider trial period range.

\begin{figure*}
  \begin{center}
    \includegraphics[scale=2.0]{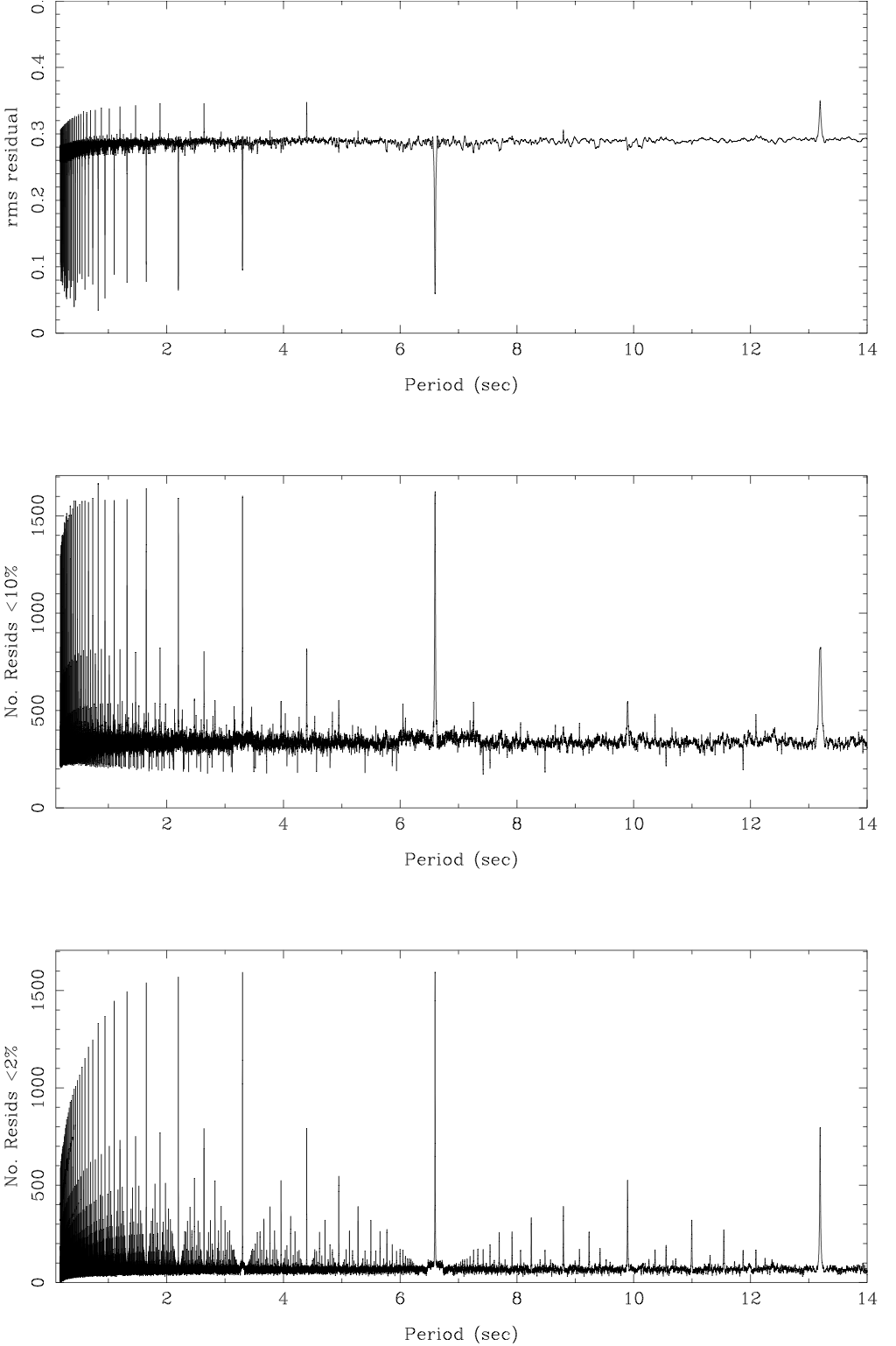} 
  \end{center}      
  \caption{\small{Trial period differences for J1841$-$14. The top
  panel shows the rms residuals for each trial period. The middle
  panel shows the number of rms residuals below 10\% and the bottom
  panel shows the number with rms residuals below 2\%.}}
  \label{fig:1841_getper}
\end{figure*}

\end{enumerate}
\begin{table*}
  \begin{center}
    \caption{The observed properties of the confirmed sources from the
    PMSingle analysis.}
    \begin{tabular}{|c|c|c|c|c|c|c|c|c|c|c|}
      \hline\hline Source & RA & DEC & DM & D & P & Epoch & w & $S_{\rm{peak}}$ & $L_{\rm{peak}}$ \\ 
      & (J2000) & (J2000) & ($\rm{cm^{-3}.pc}$) & (kpc) & (s) & (MJD) & (ms) & (mJy) & ($\rm{Jy.kpc^2}$) \\
      \hline J1047$-$58 & 10:47(1) & -58:41(7) & 69.3(3.3) & 2.3 & 1.23129(1) & 55779 & 3.7 & 630 & 3.3 \\
      J1423$-$56 & 14:23(1) & -56:47(7) & 32.9(1.1) & 1.3 & 1.42721(7) & 54557 & 4.5 & 930 & 1.5 \\
      J1514$-$59 & 15:14(1) & -59:52(7) & 171.7(0.9) & 3.1 & 1.046109(4) & 54909 & 3.3 & 830 & 7.9 \\
      J1554$-$52 & 15:54(1) & -52:10(7) & 130.8(0.3) & 4.5 &  0.12522947(7) & 54977 & 1.0 & 1400 & 28.3 \\
      J1703$-$38 & 17:03(1) & -38:12(7) & 375.0(12.0) & 5.7 & - & - & 9.0 & 160 & 5.1 \\
      J1707$-$44 & 17:07(1) & -44:12(7) & 380.0(10.0) & 6.7 & 5.763752(5) & 54999 & 12.1 & 575 & 25.8 \\
      J1724$-$35 & 17:24(1) & -35:49(7) & 554.9(9.9) & 5.7 & 1.42199(2) & 54776 & 5.9 & 180 & 5.8 \\
      J1727$-$29 & 17:27(1) & -29:59(7) & 92.8(9.4) & 1.7 & - & - & 7.2 & 160 & 0.4 \\
      J1807$-$25 & 18:07(1) & -25:55(7) & 385.0(10.0) & 7.4 & 2.76413(5) & 54987 & 4.0 & 410 & 22.4 \\
      J1841$-$14 & 18:41(1) & -14:18(7) & 19.4(1.4) & 0.8 & 6.597547(4) & 54909 & 2.6 & 1700 & 1.0 \\
      J1854+03 & 18:54:09(7) & +03:04(2) & 192.4(5.2) & 5.3 & 4.557818(6) & 54909 & 15.8 & 540 & 15.1 \\
      \hline
    \end{tabular}
    \label{tab:new_rrats}
  \end{center}
    \footnotesize{In all sources except for 1854+03 we know the
    position to the accuracy of a $1.4$-GHz beam of the 64-metre
    Parkes Telescope, which is 14 arcminutes. A more accurate position
    for J1854+03 is available from~\citet{dcm+08}. The DMs given are
    those obtained from fitting the dispersion sweep of the brightest
    pulse for each source. The distances quoted are those derived from
    the DM using the NE2001 model~\citep{cl02, cl03} of the electron
    content of the Galaxy and have typical errors of 20 percent. The
    pulse widths are those measured at 50\% intensity. The quoted
    1.4-GHz peak flux densities are determined by using the radiometer
    equation~\citep{lk05} and using the known gain and system
    temperature of the 20-cm multi-beam receiver (as given in the
    April 6, 2009 version of the Parkes Radio Telescope Users
    Guide). The typical uncertainties in this calibration are at the
    30 percent level.}
 \end{table*}
\begin{table}
  \begin{center}
    \caption{Detection statistics for the confirmed sources from the
    PMSingle analysis. $\dot{\chi}$ refers to the detected burst
    rate.}
    \begin{tabular}{|c|c|c|c|c|}
      \hline\hline Source & $N_{\rm{det}}/N_{\rm{obs}}$ & $N_{\rm{pulses}}$ & $T_{\rm{obs}}$ (hr) & $\dot{\chi}$ ($hr^{-1}$) \\
      \hline J1047$-$58 & 8/15 & 54 & 8.96 & 6.0 \\
      J1423$-$56 & 9/12 & 35 & 10.01 & 3.4 \\
      J1514$-$59 & 9/9 & 92 & 4.58 & 20.0 \\
      J1554$-$52 & 8/8 & 214 & 4.25 & 50.3 \\
      J1703$-$38 & 5/6 & 10 & 3.08 & 3.2 \\
      J1707$-$44 & 5/5 & 22 & 2.58 & 8.5 \\
      J1724$-$35 & 12/17 & 34 & 9.95 & 3.4 \\
      J1727$-$29 & 2/5 & 2 & 2.11 & 0.9 \\
      J1807$-$25 & 7/7 & 25 & 3.97 & 6.2 \\
      J1841$-$14 & 13/13 & 231 & 5.01 & 46.0 \\
      J1854+03 & 9/9 & 42 & 4.52 & 9.2 \\
      \hline
    \end{tabular}
    \label{tab:detection_statistics}
  \end{center}
 \end{table}

\section{Individual Sources}
Currently we have confirmed 11 sources, 10 of which are new. These are
listed in Table~\ref{tab:new_rrats} with some observed properties. We
note that these confirmed sources are amongst the better class 1
candidates. The detection statistics for these sources are given in
Table~\ref{tab:detection_statistics}. Coherent timing solutions have
been obtained for all the sources that have determined periods, but
spanning only a few months, so that accurate position and period
derivative determinations are not yet possible. In this section we
discuss the individual sources discovered in more detail.

\subsection*{Detection of Known Sources}
In addition to the newly discovered sources the analysis has made many
detections of previously known sources. These include detections of
300 previously known pulsars. Many of these were detected multiple
times so that there were 606 known pulsar detections in total. Up to
2006 the PMPS detected 976 pulsars (of which, at the time, 742 were
new sources, see for e.g.~\citet{lfl+06}). Since then new analyses of
results has given rise to more sources such that the ATNF pulsar
database now lists 1030 pulsars as detected in the PMPS. The 300 known
pulsar detections here then correspond to $\sim30\%$ of pulsars
detected just from single pulse searches. This is an increase on the
250 pulsars detected in the original single pulse
search~\citep{mcl_iau} with extra detections made across the entire DM range. 
%
%

%
%
The true PMSingle detection rate for known pulsar detections may even
be better than 30\% if we consider whether any of the pulsars detected
in the PMPS would actually be removed by the zero-DM filter. This
could be the case for low-DM pulsars. However, if we assume zero-DM
must remove the amplitude spectrum up to a fluctuation frequency which
is $\delta^{-1}$ harmonics in order to remove all information of the
pulsar, where $\delta=w/P$ is the pulsar duty cycle, then just 4 of
the PMPS pulsars would be removed. Assuming we need to remove just
$P/2w$ harmonics would see 16 sources (or 1.6\%) removed by the
zero-DM filter which leaves the detection rate at $\sim30\%$. In
addition some of the sources classified as candidates may turn out to
be (far side-lobe detections of) known pulsars which could potentially
boost the number of single pulse detections by a few percent.

The original 11 RRATs were all re-detected. RRAT J1819$-$1458 is
observed three times in the survey. The third PMPS detection, revealed
in this analysis, was previously unknown. This is very helpful for
timing and for attempts in connecting over a long $\sim1800$ day gap
in timing data (between survey observations and the initial single
pulse search of the data). This enables a conclusion that RRAT
J1819$-$1458 does not seem to have suffered a large glitch during this
gap in observations~\citep{glitch_paper}.

\subsection*{Discussion of Interesting Individual Sources}

J1047$-$58 was found within the same beam as the known pulsar
J1048$-$5832. This allowed for a rapid confirmation of the source by
examining archival Parkes data of the known pulsar. In the discovery
observations, all of the six detected pulses were clustered within a
$\sim100$ second window. In the followup observations of this source
there is a suggestion of such `on' times in which pulses are clustered
together in windows of up to $\sim500$ seconds. 


J1514$-$59 is detected in all observations with a period of
1.046~seconds and a large average burst rate of $\dot{\chi}\sim20$
$\rm{hr^{-1}}$. It is not seen in FFT searches of entire observations
which have all been $\sim30$ minutes in duration. However its pulses
are seen to come in approximately minute-long clumps separated by
$\sim800-1000$ seconds. Figure~\ref{fig:search_plots} shows an example
of this. Performing FFT searches focused on a small `on' region
(albeit with quite poor spectral resolution of $\sim10$~mHz) gives a
period which agrees with that obtained from examining the differences
in pulse pair arrival times. Folding the `on' regions at the nominal
period gives us a pulse profile which shows a single narrow peaked
pulse.

Analysing the intervals between the bursts shows that they do not obey
a Poission distribution. The K-S test probability that the
distribution is Poissonian is $<10^{-8}$. In fact it seems the
distribution is bi-modal with a peak at short burst intervals (of a
few periods) and another at long intervals (several hundred periods).
We find that the short intervals are consistent with a Poissonian
distribution with average expected interval of $\lambda=7$
periods. More observations are needed to determine the `peak' of the
longer interval distribution.

The long $\sim15$ minute intervals do not seem to be due to the
effects of interstellar scintillation. For this observation frequency
and DM we are in the strong scintialltion regime~\citep{lk05} and so
must consider diffractive and refractive scintillation as
possibilities. Assuming the NE2001 model~\citep{cl02,cl03} of the
Galactic free electron density we find a diffractive time-scale
$\Delta t_{\rm{DISS}}$ of $\sim30$ seconds which is much too short to
explain this modulation. Similarly the diffractive scintillation
bandwidth $\Delta f_{\rm{DISS}}$ is just $\sim10$ kHz, much
narrower than the bandwidth of a single channel in any of these
observations. Thus every channel averages many scintles and observing
this modulation is not possible. The refractive scintillation
timescale is related to the diffractive timescale by $\Delta
t_{\rm{RISS}}=(f/\Delta f_{\rm{DISS}})\Delta t_{\rm{DISS}}$ which in
this case is $\sim$10s of days which is much too long to account for
this modulation. 

It would seem then that this modulation may be something intrinsic to
the neutron star. The situation seems somewhat consistent with a
nulling pulsar where the majority of the pulses emitted during the
non-nulling phase are below our sensitivity threshold. This would
imply a nulling length of $\sim14$ minutes and a nulling cycle of
$\sim1$ minute, i.e. a nulling fraction of more than $90\%$. In
studies of 23 nulling pulsars detected in the PMPS~\citet{wmj07}
observed nulling fractions from as low as $1\%$ to as high as
$93\%$. One source showed a similar nulling cycle of $\sim515$ seconds
although most were lower. This is not inconsistent however due to the
obvious difficulty of detecting long-duration nullers. Their results
also showed large nulling fraction to be related to large
characteristic age, $\tau=P/2\dot{P}$ rather than long periods, a
relationship which can be tested for this source once a full timing
solution has been obtained. We do note the similarity between
J1514$-$59 (and J1047$-$58) and the class of ``intermittent
pulsars''~\citep{klo+06}. However the time-scales for the `on' and
`off' states in these sources can be 10s of days which is much longer
than what is seen in these RRATs. The possibility remains though that
RRATs may fit into a continuum of nulling behaviour which could range
from those sources which null for a few periods at a time at one
extreme to the intermittent pulsars at the other extreme. As the
numbers of known RRATs and intermittent pulsars increases timing
observations can be used to investigate what properties (if any,
e.g. period and age) correlate with nulling fraction.


J1554$-$52 is a strong single pulse emitter showing 35 pulses in its
discovery observation. It is also weakly detectable in FFT searches of
most observations but with much higher significance in single pulse
searches. The weakness in the FFT detection is one reason for the
previous non-detection of this source. However it is likely that this
source would have been removed by previously applied algorithms
designed to remove RFI signals. For instance frequency domain
`zapping' would have removed this source, i.e. setting certain
frequencies in the fluctuation spectrum (of a dedispersed time series)
to zero. This is done at known RFI frequencies (e.g. 50 Hz) and their
harmonics.
 With a
period of 125~ms this pulsar falls exactly into one of these zapped
regions.
The pulses are also seen to fall into two main phase windows,
reminiscent of the three such windows observed in RRAT
J1819$-$1458~\citep{glitch_paper}. 



J1724$-$35 was the first source to be discovered in this
re-processing~\citep{ekl09}. Since the 2 survey observations of this
source it has been re-observed 15 times and detected in 10 of
those. Despite having a fairly high DM its discovery is helped
immensely by the removal of very strong RFI by the zero-DM filter. In
all but 2 observations it is not detected in an FFT search. In one
observation it can be detected from focussed FFT searches of times
when strong single pulses are seen, as for J1514$-$59. In another
observation it is detected with FFT S/N of 15 which is evidence that
there is underlying weak emission in addition to the detected single
pulses. During these times when the source is detectable in
periodicity searches a folded profile can be obtained which is quite
wide and double-peaked. We note that this variation is not due to
scintillation as the scintillation time-scale of 2 seconds is too
short and the bandwidth of $\sim20$ Hz is too narrow to explain
this. The burst rate is insufficient for analysis of the intervals
between bursts at present.

J1727$-$29 is detected once in the PMPS with just one strong single
pulse. It was re-observed in a followup observation where we again
detected just one pulse. Further followup observations have not
revealed anything further from this source. This source is obviously
too weak to time and without even two pulses in a single observation a
period estimate is not possible. We expect a number of candidates like
this to be confirmed while proving impractical to continuously
monitor. Such sources will require the next-generation sensitivity of
the SKA and further serious followup efforts should be engaged when
such facilities become available.


J1841$-$14 was observed twice in the PMPS where, as we can see from
Figure~\ref{fig:lowDM_blindness}, its detection was hindered by the
presence of strong RFI in the initial analysis. The source has been
re-observed 11 times and detected in all cases. It has the lowest DM
of any RRAT found in the PMPS (and lower than $95\%$ of normal radio
pulsars). It has a very high burst rate of $\dot{\chi}\sim40$
$\rm{hr^{-1}}$ but it is undetectable in an FFT search. However this
seems to be due to the insensitivity of FFT searches in detecting long
periods\footnote{This is due to red-noise and in the case of the
zero-DM filter a suppressed fluctuation spectrum at low
frequencies}. Using a Fast Folding Algorithm~\citep{kml+09}, an
alternative periodicity search, more sensitive than FFTs for high
periods, we detect the source at the correct period. The average FFA
S/N is 9 as compared to the peak single pulse S/N of $\sim60$ and many
pulses per observation with single pulse S/N $\geq15$. It has a very
long period of 6.596 seconds as determined from factorising the pulse
pair time-of-arrival differences. Figure~\ref{fig:1841_getper} shows
the results of this period determination. Folding the observations at
this period shows a narrow pulse profile. 
The pulses observed are among the brightest seen for RRATs with
typical peak flux densities (at 1.4 GHz) of $\sim1$ Jy and a maximum
peak flux density observed of $1.7$ Jy. While most of the pulses are
narrow at $\sim2$ ms there are few pulses detected with pulse widths
of as wide as 20~ms. The high burst rate means that obtaining a
sufficiently large number of TOAs at regular intervals to obtain a
coherent timing solution should be straightforward. Obtaining an
accurate timing position will be useful for a detection of this source
at X-rays which appears very promising as the source is nearby at a
distance of $\sim800$ pc.

J1854+03 was observed once in the PMPS and has since been re-observed
eight times and detected in all cases. This source is one previously
identified by the 1.4-GHz PALFA survey~\citep{dcm+08}. As the PALFA
position is much more accurate than that which we were able to obtain
with the Parkes telescope (due to the much smaller beam size of the
Arecibo telescope) it may be possible to determine a period derivative
for this source on a shorter timescale than for the other
sources. This is because, typically, determining a period derivative
takes a year of timing observations so that the effects of positional
uncertainty (which shows year-long sinusoidal patterns in timing
residuals) and the slow-down rate of the star can be
disentangled. This source has a high burst rate which is
$\dot{\chi}\sim10$ $\rm{hr^{-1}}$. It is undetectable in FFT searches
and has a long period of 4.558 seconds. However this more distant
source (it is $\sim6$ times further away than J1841$-$14) shows weak
pulses. Typical peak flux densities (at 1.4 GHz) are $\sim100$ mJy but
the brightest observed pulse is $\sim540$ mJy. The pulse widths are
typically $\sim15$~ms and there are no indications of clumps of
emission on which to focus FFT searches.

\section{Discussion}
The motivation for this re-processing of the PMPS is to find more RRAT
sources. The reason for this is that the RRATs seem to represent a
very large population of Galactic neutron stars, likely larger than
the regular radio pulsar population. It is therefore important to
clearly describe differences between RRATs and pulsars, leading to a
meaningful definition of RRATs. Having discovered the existence of
RRATs by their bursty emission, our on-going studies of their emission
properties suggests that a useful definition will go beyond a simple
description of this burst behaviour.
%

\subsection*{What is a RRAT?}
From a detection point of view it is difficult to define a RRAT. One
possible definition is that of a source only detectable in single
pulse searches and not in periodicity searches. As the RRATs seem to
have intrinsically longer periods one might think this is a selection
effect. For equal amplitude distributions one might indeed expect less
periodicity search detections as the number of pulse periods in a
given observation time is less. However McLaughlin et al. (2009b) show
that the period distributions are different with high significance and
that the single pulse search sensitivity does not select against
low-period RRATs should they exist. It seems therefore that the RRATs
have intrinsically longer periods than most radio pulsars.

One might extend the definition to be that of a source more easily
detected in single pulse searches than periodicity searches. The
so-called ``intermittency ratio'', R, is defined as the ratio of the
single pulse and periodicity search S/N ratios. Thus a RRAT would be
considered to be a source that has $R>1$ but this immediately has
several problems as the intermittency ratio varies between
observations and some kind of arbitrarily averaged value of $R$ would
be needed. Also, as far as detectability is concerned several single
pulses with, say, S/N of 10 are much more easily detectable than a
periodicity search with a similar S/N of 10 due to the different
false-positive noise levels for each method. Thus the same value of
$R$ can describe very different scenarios and its usefullness as a
measure of detectability is limited. Another problem is that our
current data suggest that a RRAT detected at 1.4 GHz can behave
differently at a different observing frequency. Corresponding studies
of the yet unknown spectra through multi-frequency observations of
RRATs are underway (Keane et al. in prep; Miller et al. in prep.).

A detailed definition of RRATs therefore needs to include all of this
information - intermittency, amplitude and spectral distributions,
multi-frequency behaviour as well as period and period derivative
properties and the derived quantities related to these in
particular. For example, the inferred magnetic fields distribution in
RRATs has been shown to be different to that of the pulsars with a
high significance (McLaughlin et al. 2009b). In addition one source
has also shown anomolous glitch activity (Lyne et al. 2009) although
it is unknown whether this is characteristic of the entire population
of RRATs. In Table~\ref{tab:properties} we summarise the observed
properties of pulsars, magnetars, and the RRATs. The INSs are also
likely to fit into the overal picture of how these neutron star
classes are related but are not included here for lack of information
on their radio properties. Further observations will constrain the
ranges of these properties further for RRATs which will allow us to
refine the classification of RRAT sources.

\begin{table*}
  \begin{center}
    \caption{Observed characteristics of (non-recyled) pulsars,
    magnetars and RRATs. Information in this table is taken from many
    sources: in addition to those references mentioned in the main
    text these
    include~\citet{crh+06,crj+07,crj+08,kwv+09,ljk+08,mkk+00,timing_paper,ssw+09}
    and the McGill Magnetar Catalogue
    http://www.physics.mcgill.ca/$\sim$pulsar/magnetar/main.html.
    $B_{\rm{LC}}$ and $B_{\rm{S}}$ refer to the inferred magnetic
    fields at the light cylinder and stellar surface respectively
    assuming a star with a dipolar magnetic field (see
    e.g.~\citet{lk05}. $B_{\rm{quantum}}$ is the quantum-critical
    magnetic field above which the separation of Landau levels for
    synchrotron orbits exceeds the electron rest mass. The quantities
    L, V and I refer to linear polarisation, circular polarisation and
    total intensity respectively. We note that although there are 29
    published RRAT sources there are determined periods for just 27
    and spin-down properties like $\dot{P}$, $B_{\rm{S}}$, $\dot{E}$
    and $\tau$ are known for just 7 sources~\citep{timing_paper}.}
    \begin{tabular}{|l|l|l|l|}
      \hline\hline Property & Pulsar ($\sim1700$ sources) & Magnetar (18 sources) & RRAT (29 sources) \\
      \hline 
      Period (s) [range,median] & $\sim0.03-8.5$, $0.54$ & $2.1-11.8$, $7.3$ & $0.1-6.7$, $2.1$ \\
      $\log$(Period derivative) & $-17$ to $-12$ & $-13$ to $-9$ (14 measured) & $-15$ to $-13$ (7 measured) \\
      $\log(\dot{E})$ (erg/s) & $30-38$ & $31-35$ & $31-33$ \\
      $\log(B_{\rm{S}})$ (gauss) [range,median] & $10.0-14.0$, $12.03$ & $13.0-15.0$, $14.5$ $>B_{\rm{Quantum}}$ limit & $12.4-13.7$, $13.0$ \\
      $\log(B_{\rm{LC}})$ (gauss) & $-2$ to $6$ & $\sim-0.01$ to $1.1$ & $\sim-0.1$ to $1.7$ \\
      $\log(\tau=\dot{P}/2P)$ (yr) & $3-9$ & $2-6$ & $5-6$ \\
      Observed Radio on-fraction & $0.05-1$ & $0-1$ & $10^{-3}-10^{-2}$ \\
      Transient behaviour & nulls, intermittent PSRs & X-ray and radio variability & $\dot{\chi}\sim1$ $\rm{min}^{-1}$ to $\sim1$ $\rm{hr}^{-1}$ \\
      Bursting behaviour & GRPs, giant micro-pulses, & X-ray and soft $\gamma$-ray bursts & $\lesssim10$ Jy single pulses \\ & nano-giant pulses \\
      No. of components/subpulses & $\geq1$ components & large phase range, many  & $\geq1$ components \\ & & components in XTE 1810$-$197 & $\geq1$ \\
      Beaming fraction ($f_{\rm{beam}}(P)$) & $0.09[\log(P/s)-1]^2+0.03$ & unknown & unknown \\
      Amplitude distribution & GRP: gaussian + power & power-law or log-normal, & some log-normal (see \S 4), \\ & law, indices $-1.5$ to $-4$ & depending on which component  & some power law, index $-1$\\
      Radio spectra & $\nu^{-\alpha}$, $\langle\alpha\rangle=1.8\pm0.2$ & flat but variable & unknown \\
      X-ray spectra & thermal and/or power-law & thermal + power law & thermal in J1819$-$1458 \\
      Polarisation & low $-$ high & $\lesssim100$ percent & $L/I\sim0.37$, $V/I\sim0.06$ \\ & & & for J1819$-$1458 \\
      \hline
    \end{tabular}
    \label{tab:properties}
  \end{center}
 \end{table*}

\subsection*{Distant Pulsars?}
Closely related to the question of the nature of a RRATs are of course
models that explain RRATs as distant analogues of pulsars with a pulse
amplitude distribution with a long tail. ~\citet{wsrw06} have shown
that the nearby PSR B0656+14 (at a distance of 288 pc) would appear
RRAT-like if moved to typical RRAT distances. The amplitude
distribution of the pulses from this source are log-normal but can be
described by a power law with index of between $-2$ and $-3$ at the
high flux density end. We can test this scenario if we assume that
RRATs emit pulses according to a power law amplitude probability
distribution like
\begin{equation}
  P \propto S^{-\alpha}
\end{equation}
between flux density values $S_{\rm{min}}$ and $S_{\rm{max}}$. We
detect all pulses above $S_{\rm{thresh}}$ where $S_{\rm{min}}\leq
S_{\rm{thresh}}\leq S_{\rm{max}}$. This means that the fraction of
observable pulses, g, is given by
\begin{equation}
  g=\frac{\int_{S_{\rm{thresh}}}^{S_{\rm{max}}}P(S)dS}{\int_{S_{\rm{min}}}^{S_{\rm{max}}}P(S)dS}
\end{equation}
where $S_{\rm{max}}$, $S_{\rm{thresh}}$ and g are all known. The
observed values for g for the 9 new sources with known
periods\footnote{To determine g the period must be known as g is the
average number of pulses per period, or
$g=(\dot{\chi}/hr^{-1})(P/s*3600)$} are given in
Table~\ref{tab:power_laws}. Thus for various chosen power law indices
$\alpha$, we can determine $S_{\rm{min}}$. This can be used to
determine the distance where we would need to move the RRAT to see all
of its pulses (i.e. observe it like a pulsar which emits continuously
with $g=1$) from $D_{\rm{new}}=D(S_{\rm{min}}/S_{\rm{thresh}})^{1/2}$
by noting $L=SD^2$ where $L$, the radio luminosity\footnote{$L$ has units
of $\rm{Jy.kpc^2}$ or equivalently $\rm{W.Hz^{-1}}$}, is the more
intrinsic quantity.

\begin{table*}
  \begin{center}
    \caption{The required distances to see continuous emission if
    these sources emit according to various power laws.}
    \begin{tabular}{|c|c|c|c|c|c|c|}
      \hline\hline Source & g & D (kpc) & & & $g=1$ distances (kpc) \\
       & & & ($\alpha=1.5$) & ($\alpha=2$) & ($\alpha=3$) & ($\alpha=4$) \\
      \hline J1047$-$58 & 0.0021(1/476) & 2.3 & 0.01 & 0.12 & 0.50 & 0.83 \\
      J1423$-$56 & 0.0014(1/714) & 1.3 & $<0.1$ & 0.05 & 0.25 & 0.43 \\
      J1514$-$59 & 0.0058(1/172) & 3.1 & 0.03 & 0.26 & 0.86 & 1.31 \\
      J1554$-$52 & 0.0016(1/625) & 4.5 & 0.01 & 0.20 & 0.91 & 1.54 \\
      J1703$-$38 & - & 5.7 & - & - & - & - \\
      J1707$-$44 & 0.0137(1/73) & 6.7 & 0.14 & 0.84 & 2.30 & 3.28 \\
      J1724$-$35 & 0.0013(1/769) & 5.7 & 0.03 & 0.30 & 1.18 & 1.94 \\
      J1727$-$29 & - & 1.7 & - & - & - & - \\
      J1807$-$25 & 0.0048(1/208) & 7.4 & 0.61 & 2.00 & 3.05 \\
      J1841$-$14 & 0.0845(1/11.8) & 0.8 & 0.10 & 0.24 & 0.43 & 0.53 \\
      J1854+03 & 0.0118(1/85) & 5.3 & 0.10 & 0.61 & 1.75 & 2.53 \\
      \hline
    \end{tabular}
    \label{tab:power_laws}
  \end{center}
 \end{table*}

Table~\ref{tab:power_laws} shows the distances one would need to move
the sources discussed here to see continuous emission if they would be
regular pulsars with amplitude distributions that follow
power-laws. We can see that for the steepest power laws the source is
not required to move very much nearer for all pulses to become
visible. As the power law index gets shallower the source must be
brought ever closer to be seen as a continuous emitter. For
$\alpha\lesssim1.5$ the change in distance becomes unreasonable such
that if all RRATs were continously emitting pulses with energy
distributions as a power law with $\alpha\lesssim1.5$, almost none of
these sources would ever appear as continuous emitters for a
reasonable distance distribution and such sources would be seen as
distinct from pulsars. We thus conclude that RRAT emission could be
explained as coming from distant pulsars, i.e. continuous emitters,
with steep power-law distributions only. For shallower pulse
distributions a power-law alone cannot explain the observed RRAT
emission as being due to distant pulsars. However the sources may
still be seen as continuous if the distribution were to break to,
e.g., a log-normal distribution at low flux densities. We can compare
these results with the amplitude distributions from the initial RRAT
discovery paper which showed some distributions being consistent with
power law indices of $\alpha=1$~\citep{mll+06}. Clearly, the amplitude
distribution of pulses will provide a powerful discriminator between
sources that can be explained as distant pulsars and those which
cannot.

\begin{figure}  
  \begin{center} 
    \includegraphics[trim = 20mm 20mm 10mm 20mm, clip, scale=0.35]{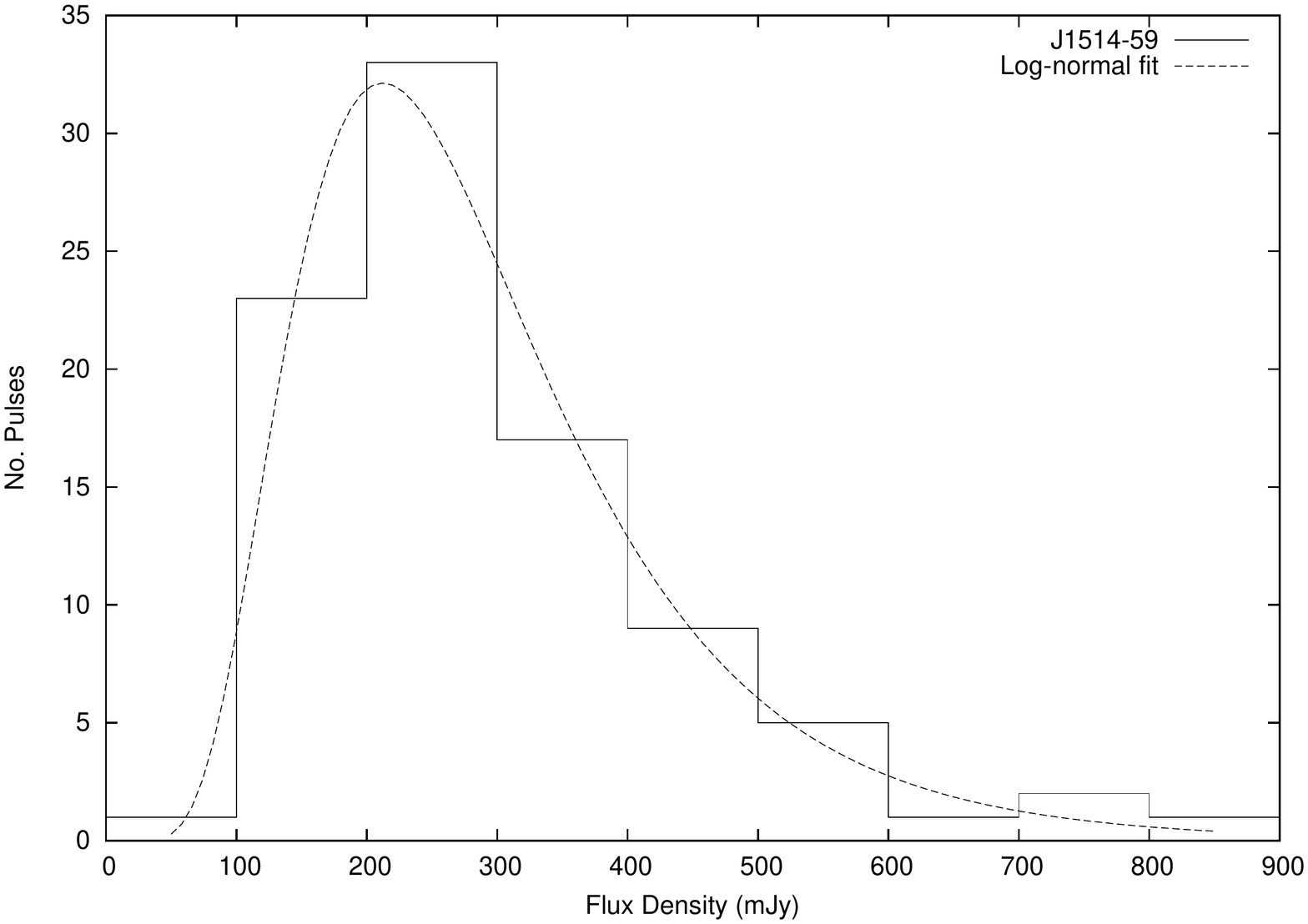}
    \includegraphics[trim = 20mm 20mm 10mm 20mm, clip,scale=0.35]{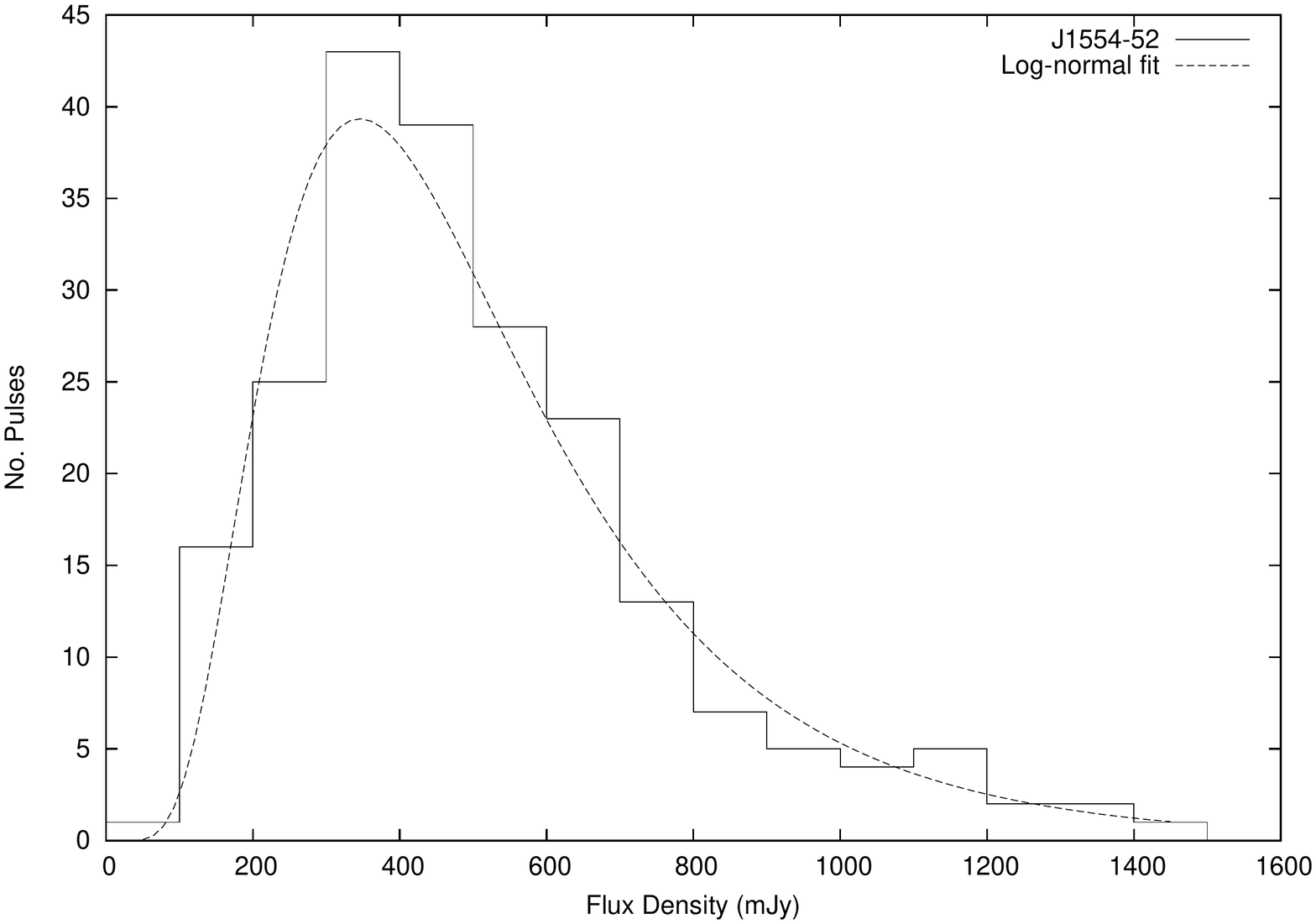}
    \includegraphics[trim = 20mm 20mm 10mm 20mm, clip,scale=0.35]{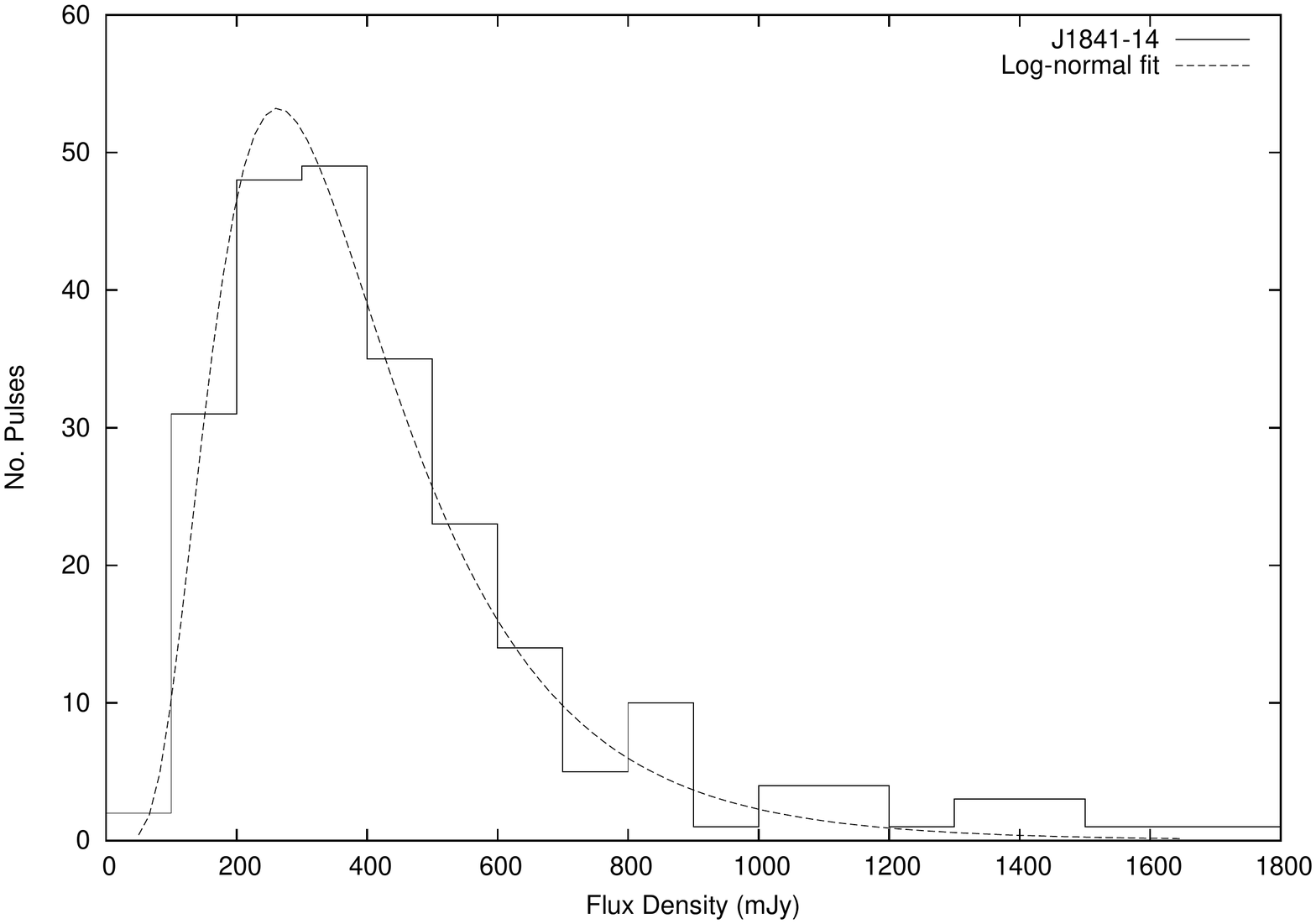}
  \end{center}      
  \caption{\small{Amplitude distributions for: (Top) J1514-59;
  (Middle) J1554-52 and; (Bottom) J1841-14.}}
  \label{fig:amp_dist}
\end{figure}

Figure~\ref{fig:amp_dist} shows the amplitude distributions for
J1514$-$59, J1554$-$52 and J1841$-14$, the three sources discussed
here with the highest number of detected pulses. These distributions
are found not to be consistent with a power-law distribution but
instead are well fitted by log-normal distributions, the parameters of
which are given in Table~\ref{tab:lognormal_fit}. The best-fit curves
are over-plotted on the observed distributions in
Figure~\ref{fig:amp_dist}. For these three sources there is a low flux
density turn-over. It is not clear whether this is an intrinsic
turn-over or simply due to the sensitivity threshold. The flux density
threshold for a single pulse depends on the pulse width. Plugging in
the known observing parameters for Parkes into the radiometer equation
gives a single pulse peak flux of: $S_{\rm{peak}}\approx245$
$\rm{mJy}(w/\rm{ms})^{-\frac{1}{2}}$ assuming a 5-sigma detection
threshold. Although the widths of the pulses vary from pulse to pulse
we can take the average widths from Table 2 to get sensitivity
estimates of 135 mJy, 250 mJy and 150 mJy for J1514$-$59, J1554$-$52
and J1841$-$14 respectively. If the turn-overs were intrinsic to the
sources then it would suggest that we are not just seeing the
brightest pulses from a continuously emitting source but rather that
we are seeing most pulses which are emitted. If this is the case, the
bursty behaviour is indeed due to the lack of continuous emission and
an innate property of the sources. For the remaining sources the
number of pulses detected is as yet still too low for such an
analysis. Continued observations will allow accurate determination of
amplitude distributions of all the sources as more observations are
made.

\begin{table}
  \begin{center}
    \caption{The best-fit parameters to the amplitude distributions in
    Figure~\ref{fig:amp_dist} for a log-normal probability density
    distribution of the form:
    $P(x)=(a/x)\rm{exp}[-\frac{(\rm{ln}x-b)^2}{2c^2}]$. The parameter
    a is an arbitrary scaling factor and the values given here
    correspond to the scales used in Figure~\ref{fig:amp_dist}.}
    \begin{tabular}{|c|c|c|c|}
      \hline\hline Source & a & b & c \\
      \hline J1514$-$14 & 7613(748) & 5.57(0.03) & 0.47(0.03) \\
      J1554$-$52 & 15662(1182) & 6.13(0.03) & 0.53(0.03) \\
      J1841$-$14 & 16130(1704) & 5.86(0.04) & 0.53(0.04) \\
      \hline
    \end{tabular}
    \label{tab:lognormal_fit}
  \end{center}
 \end{table}

\subsection*{The Emerging RRAT Population}

We note that the scope of any definition can and should be wary of
those radio pulsars which are sufficiently weak and/or distant not to
be detected by periodicity searches. Such sources are automatically
accounted for by the predicted Galactic pulsar population. For
instance there are estimated to be $\sim150,000$ radio pulsars in our
Galaxy with a radio luminosity $L=SD^2>0.1$
$\rm{mJy.kpc}^2$~\citep{lfl+06}, or $\sim100,000$ with $L>1.0$
$\rm{mJy.kpc}^2$~\citep{vml+04}. These estimates are arrived at from
accounting for known observational selection effects of surveys and
extrapolating from the observed properties of pulsars (e.g. the
observed luminosity distribution and empirical beaming
models). However we note that if the population of RRATs is indeed a
factor of a few times larger than that of the radio pulsars not all
RRATs can be accounted for as weak and/or distant pulsars. With the
addition of these 11 new PMPS RRAT sources there are now 22 confirmed
RRAT sources in the survey. There are a further 6 single pulse sources
in the literature recently reported by~\citep{dcm+08} (although some
of these are detectable in FFT searches also) and one source
discovered in the GBT350 survey~\citep{hrk+07}. Of these 29 currently
known sources 27 have determined periods and seven (of the original
11) sources now have known period
derivatives~\citep{timing_paper}. The initial RRAT population estimate
suggested that there were $\sim4$ times as many RRATs as
pulsars~\citep{mll+06}. We defer a complete population analysis until
we have investigated all of our best candidates but we can already
comment on some of the selection effects impacting upon this
estimate. These include the RRAT beaming fraction, their observed
on-off fraction and the fraction missed due to contaminating RFI: (a)
With no reliable information on a beaming model for RRATs, we continue
to use the empirical pulsar beaming model,
$f_{\rm{beam}}(P)=0.09[\log(P/s)-1]^2+0.03$, of~\citet{tm98}. Still,
it remains to be seen what the actual RRAT beaming model might be or
whether this beaming model even still applies for pulsars with long
periods. The latter is highlighted by the recent discovery of a slow
pulsar with an extremely narrow pulse by Keith et al.~(2008). If
beaming fractions were in fact overestimated, the projected
populations of both pulsars and RRATs may be much larger. (b) The
assumed averaged on-off fraction during a PMPS observation was taken
to be 1/2, i.e. half of the RRATs that exist were assumed not to emit
pulses during the 35~min observation which was consistent with the
observed burst rates at the time. With the discovery and continued
monitoring of a growing population of known RRATs it will be possible
to determine the true burst rate distribution and thus to improve this
value in population estimates. (c) The factor used to compensate for
those sources missed due to the presence of RFI was quite uncertain
and taken to be 0.5. Taking into consideration the pointings with
strong residual RFI and the fraction of sources removed by zero-DM in
the light of the PMSingle analysis the number of beams still affected
by RFI is $(41561-693)*0.016+693$ which is $\sim3\%$ of all beams, so
that 97\% are now cleaned of RFI. We have doubled the number of known
RRATs while simultaneously reducing the fraction missed by RFI to
almost zero. The inferred population estimate is thus related to the
initial estimate by a factor of $(97/50)/2\approx1$ so that the
confirmed number of new sources is consistent with the original
population estimate~\citep{mll+06}. However, if even a small fraction
of the large number of other candidates are confirmed as new RRATs,
the initial population estimate will need to be revised upwards so
that the number of transient radio emitting neutron stars in the
Galaxy is even larger than initially thought. Of course with the
continued monitoring and thus characterisation of more sources we can
also improve our knowledge of the other selection effects.

Assuming the population of RRATs is indeed much larger than that of
the normal radio pulsars, the neutron star birthrate problem
remains. In addition to pulsars and RRATs there are the INSs,
magnetars, neutron stars in accreting binary systems and the Central
Compact Objects (of which some are thought to be neutron stars,
e.g. see~\citet{d08}). ~\citet{kml+09} suggest that more than 30 new
INS sources need to be discovered before unfavourable beaming can be
ruled out, i.e. that these sources are pulsars and/or RRATs whose
beams do not cross our line of sight. Recently several tens of new INS
candidates have been identified~\citep{pmj+09} so that progress may
come in this area despite the difficulty in finding such
sources~\citep{pph+08}. Besides pulsars many more sources are needed
for the other classes to reach a conclusive answer as to the links (if
any) between these populations.


\section{Conclusions}
We have presented an overview of the steps involved in performing
searches for sources showing single pulses of radio emission. These
have enabled us to discover 11 new PMPS sources from which between 2
and 231 pulses have been detected. Underlying periodicities for these
sources lie in the range of 125~ms up to 6.6~s. These join the
original 11 RRAT sources identified previously in the survey. We have
reduced the uncertainties with regard to the number of sources missed
due to the contaminating effects of RFI. The projected population of
these sources still appears to be larger than the regular radio
pulsars. However we are yet to determine the burst rate and beaming
distributions for the RRATs - key ingredients in a complete population
synthesis. These characteristics will be determined with continued
monitoring of the new sources and followup investigations of the other
promising candidates (of which there are now more than 100). It seems
also that there will be many candidates for whom it will be
impractical or impossible to follow up at present with current
observing facilities. These will require followup with instruments
like LOFAR, FAST or the SKA. We note that these these instruments will
produce extraordinarily large volumes of data so that searching for
transient RRAT-like sources will necessitate the development of
automated algorithms which will use the steps as outlined above.

\section*{Acknowledgments}
We would like to thank the anonymous referees for providing helpful
comments which have improved the quality of this work.  EK
acknowledges the support of a Marie-Curie EST Fellowship with the FP6
Network ``ESTRELA'' under contract number MEST-CT-2005-19669. MAM is
supported by a WV EPSCoR grant.




\end{document}